\documentclass[twocolumn]{IEEEtran}
\usepackage{multicol}
\usepackage{multirow}
\newcommand{\keywords}[1]{\par\textbf{Keywords:} #1}

\usepackage{url} 
\usepackage[utf8]{inputenc} 
\usepackage{booktabs} 
\usepackage{graphicx}

\usepackage{lipsum}
\usepackage[T1]{fontenc}
\usepackage[sort]{natbib}
\usepackage{hyperref}
\usepackage{balance}

\title{Technical Report: Evaluation of ChatGPT Model for Vulnerability Detection}
\author{Anton Cheshkov, Pavel Zadorozhny, Rodion Levichev\\}
\date{\today}
\begin{document}

\maketitle

\begin{abstract}
In this technical report, we evaluated the performance of the ChatGPT and GPT-3 models for the task of vulnerability detection in code. Our evaluation was conducted on our real-world dataset, using binary and multi-label classification tasks on CWE vulnerabilities. We decided to evaluate the model because it has shown good performance on other code-based tasks, such as solving programming challenges and understanding code at a high level. However, we found that the ChatGPT model performed no better than a dummy classifier for both binary and multi-label classification tasks for code vulnerability detection.
\end{abstract}


\section{Introduction}
This technical report evaluates the performance of the ChatGPT and GPT-3 models for the task of Java code vulnerability detection. The report presents a preliminary study based on our dataset from GitHub of patch-level Java code containing both vulnerable and patched files. We used the API provided by OpenAI to send requests to GPT models for vulnerability detection. The purpose of this report is to answer the question of whether large language models such as ChatGPT and GPT-3 can solve the problem of vulnerability detection. The report includes a detailed description of the dataset collection, data preprocessing, and evaluation methodology. Finally, we provide an analysis of the results and recommendations for future research.

\section{Problem Definition}
In previous studies, GPT has been demonstrated to possess the ability to understand and generate code. The model's proficiency in coding has been tested on various coding challenges, such as HumanEval and LeetCode, where it achieved remarkable results, outperforming other LLMs (Large Language Models) and being comparable to human performance (\cite{bubeck2023sparks}). Given these findings, we aim to investigate whether the same model can also be employed for other tasks, such as vulnerability detection.

The purpose of this report is to answer the question: \textbf{can large language models such as ChatGPT solve the problem of vulnerability detection?}

\section{Related work}
In recent years, there have been several studies conducted on the evaluation and capabilities of large language models for code generation and analysis. Some of the studies that have informed our research include:

In the technical report "Sparks of Artificial General Intelligence: Early experiments with GPT-4" \cite{bubeck2023sparks}, the authors compared the capabilities of GPT-4, GPT-3, and GPT-3-5 in understanding and generating code. The report demonstrated the proficiency of these models in various coding challenges, such as HumanEval and LeetCode, where they achieved remarkable results that were comparable to human performance. The models have also shown their ability to write code from high-level instructions, as demonstrated by the zero-shot creation of a 3D game using HTML and JavaScript, as well as the implementation of a custom optimizer module for deep learning applications.

In addition to code generation, the models exhibited a high level of understanding of existing code, written by others. This includes the ability to reason about code execution, as evidenced by their correct explanation of the output of a C program involving the size of two structures. Furthermore, the models demonstrated the ability to execute non-trivial Python code, simulating the code in natural language while keeping track of several variables, nested loops, dictionaries, and recursion.

The GPT series has also been evaluated on various tasks such as summarization, sentiment analysis, question answering, among others, and has shown outstanding performance on these tasks (\cite{brown2020language, yoo2021gpt3mix}).

In the study "Evaluating Large Language Models Trained on Code" (\cite{chen2021evaluating}), the authors investigated the performance of large language models, such as GPT-3, when trained on code. They explored the model's ability to understand and generate code, offering valuable insights into the potential of using such models for code-related tasks. According to their experiments, Codex demonstrates proficiency in generating certain types of code components but struggles with others, such as SQL and shell injection payloads. The study finds that Codex underperforms compared to rudimentary Static Application Security Testing (SAST) tools in vulnerability discovery. As model capabilities improve, further research is necessary to understand the potential for large language models to excel at discovering high-dimension vulnerabilities, such as "business logic" vulnerabilities.

In "A Comprehensive Capability Analysis of GPT-3 and GPT-3.5 Series Models" (\cite{ye2023comprehensive}), the authors provided a detailed analysis of the capabilities of GPT-3 and GPT-3.5 series models. They covered various tasks, including summarization, sentiment analysis, and question-answering, presenting an extensive understanding of the models' strengths and weaknesses.

\section{Dataset Collection}
\begin{figure}[h]
    \centering
    \includegraphics[width=0.5\textwidth]{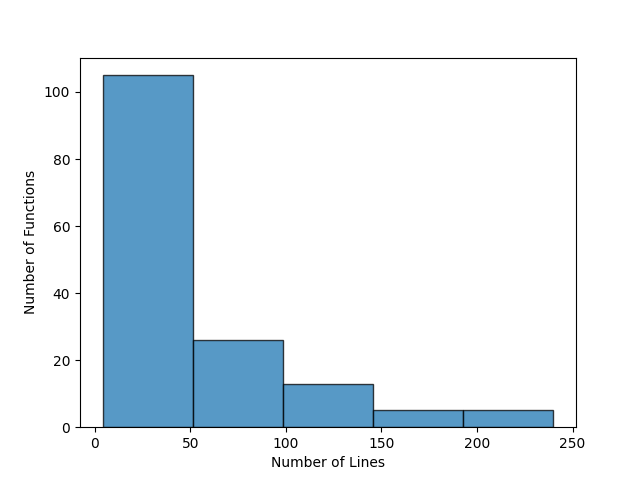}
    \caption{Function size distribution in the dataset.}
    \label{fig:histogram}
\end{figure}

We collected a dataset of Java files containing both vulnerable and patched code from open GitHub repositories, 
which were sourced based on their commit history. The vulnerable code was selected manually by domain experts and 
was known to contain vulnerabilities. The patched code was obtained by applying a patch to the vulnerable code. 
The distribution of CWE types of vulnerabilities is given in the Table~\ref{appx:distribution_CWE} of appendix. 
The table below represents only a portion of the repositories used in our dataset, and a sample of the projects and the number of 
vulnerabilities is provided in the appendix Table~\ref{appx:project_count}. The appendix Figure~\ref{fig:histogram} shows 
function size distribution of the collected dataset.

\section{Data Preprocessing}
\subsection{Binary Classification}
We extracted a subset of files from the dataset, specifically those in which a single function was modified. We then introduced a function containing a known vulnerability and its associated patched version into the subset, resulting in a dataset consisting of a total of 308 samples.
\subsection{Multi-label Classification}
The dataset consists of a certain number of Common Weakness Enumeration (CWE) vulnerability types, which are used for multi-label classification. From this dataset, the top five vulnerabilities (CWE-20, CWE-200, CWE-502, CWE-611, CWE-79) were selected based on their frequency, resulting in a total of 60 samples with vulnerabilities and 60 patched samples of these functions (Negative). We used the list of prompts presented in Table~\ref{appx:prompts} of appendix.

\section{Using Models API }
In the study, two language models were used: the Davinci model, which is considered one of the best and most advanced models within the GPT-3 family, and the GPT-3.5-turbo model, which is a specific implementation of the GPT-3 architecture designed for chatbot applications. We used the API provided by OpenAI to send requests to GPT models for each Java function extracted during the data preprocessing step. These prompts were designed to ask whether the Java code contained a vulnerability, with the goal of generating a "Yes" or "No" response from both GPT-3 and ChatGPT. It should be noted that the ChatGPT API may differ from the model provided in the web version.

\section{Evaluating Methodology}
In our evaluation methodology, we performed the following actions: For binary classification, we collected a dataset containing 308 samples, including vulnerable and patched Java functions. For multi-label classification, we compiled a dataset of 120 samples, with 60 vulnerable functions and 60 patched samples, focusing on the top five Common Weakness Enumeration (CWE) vulnerabilities. We then queried the models using the API with appropriate prompts. After obtaining the results, we calculated relevant metrics and compared the models' performance to a baseline model.
\subsection{Metrics}
To evaluate the performance of both GPT-3 and ChatGPT models for vulnerability detection, we used the following metrics:
\begin{itemize}
\item Accuracy: The percentage of correctly classified samples out of the total samples.
\item Precision: The percentage of true positives among the predicted positive samples.
\item Recall: The percentage of true positives among the actual positive samples.
\item F1 Score: The harmonic mean of precision and recall, providing a balanced metric to evaluate the model's performance.
\item AUC: A measure of the performance of a binary classification model, represented by the area under the Receiver Operating Characteristic (ROC) curve.

\end{itemize}
\subsection{Adjusting Model Parameters}
In the evaluation process, we modified the temperature parameter in GPT-3 and ChatGPT models to decrease output randomness, while keeping other parameters at their default values. Although the temperature had a minimal effect on the responses, we set it to 0 to ensure more focused and deterministic outputs.

\subsection{Model Comparison}
We compared three models, namely GPT-3, ChatGPT, and a baseline model, across binary and multi-label classification tasks to evaluate their performance in detecting vulnerabilities in Java code.

\subsection{Baseline Model}
In order to provide context for the results, we compared GPT-3 and ChatGPT's performance to a baseline dummy classifier, 
which draws labels according to its distribution in the dataset. 

We averaged the results of 1,000 runs of the dummy classifier. We do this 
to mitigate random fluctuations due to the relatevely small size of the dataset and obtain a more accurate performance estimate.

\section{Evaluation Results}
\subsection{Binary classification}

\begin{table} 
\centering
\begin{tabular}{l c c c c}
\toprule
& \multicolumn{4}{c}{Binary} \\ 
\cmidrule(l){2-5} 
Model & Precision & Recall &  F1-Score & AUC \\ 
\midrule 
text-davinci-003 & 0.50 & 0.99 & 0.67 & 0.51  \\ 
gpt-3.5-turbo & 0.51 & 0.80 & 0.62 & 0.51 \\ 
baseline & 0.50 & 0.50 & 0.50 & 0.50 \\ 
\bottomrule 
\end{tabular}
\smallskip 
\caption{Comparing 'text-davinci-003', 'gpt-3.5-turbo' and baseline models on binary classification task with 308 samples.} 
\label{tab:binary}

\end{table}

Initially, we evaluated the model's performance on a binary classification task, the dataset consisting of a total of 308 samples. 
The Table~\ref{tab:binary} compares the performance of three models on a binary classification task using 308 samples. The models evaluated are 'text-davinci-003', 'gpt-3.5-turbo', and baseline. The evaluation metrics used are precision, recall, F1-score, and AUC.

Despite GPT models have higher recall and F1 scores compared to the baseline model\footnote{The constant dummy classifer shows F1 
score about $0.67$ for the balanced sample}, the AUC score is not better than random guess model.

This observation indicates that GPT models are not able to distinct vulnerable code and its fixed non-vulnerable version. This models 
tend to label examples as positive that leads to a high false positive rate.

Also we used different prompts to ensure that our results were not influenced by a specific prompt. 
List of Vulnerability Detection Prompts for binary classification task:
\begin{itemize}
\item write whether the vulnerability is contained in the code in the Yes/no format:
\item does the following function contain a vulnerability write in Yes/no format and why?:
\item tell whether this function is vulnerable in the Yes/no format:
\item write whether this function is vulnerable:
\end{itemize}
We conducted an experiment using different prompts, and our findings indicate that the scores remain consistent across the various prompts, 
showing no significant difference in performance.

\subsection{Multi-label classification}

\begin{table} 
\centering
\begin{tabular}{l c c c c}
\toprule
& \multicolumn{4}{c}{Multi} \\ 
\cmidrule(l){2-5} 
CWE type & Precision & Recall &  F1-Score & support \\ 
\midrule 
CWE-20 & 0.333 & 0.071 & 0.118 & 14 \\
CWE-200 & 0.000 & 0.000 & 0.000 & 9 \\
CWE-502 & 0.200 & 0.143 & 0.167 & 7 \\
CWE-611 & 0.083 & 0.077 & 0.080 & 13 \\
CWE-79 & 0.333 & 0.176 & 0.231 & 17 \\
Negative & 0.440 & 0.667 & 0.530 & 60 \\
\midrule 
\midrule 
accuracy   &  & & 0.383 & 120 \\
macro avg & 0.232& 0.189    & 0.187 & 120 \\ 
weighted avg & 0.327& 0.383    & 0.330 & 120 \\
\bottomrule 
\end{tabular}
\smallskip 
\caption{Performance Evaluation of the 'gpt-3.5-turbo' Model on 5 CWE Types Using 120 Samples} 
\label{tab:gpt3_5_multi}
\end{table}

\begin{table} 
\centering
\begin{tabular}{l c c c c}
\toprule
& \multicolumn{4}{c}{Multi} \\ 
\cmidrule(l){2-5} 
CWE type & Precision & Recall &  F1-Score & support \\ 
\midrule 
CWE-20 & 0.333 & 0.071 & 0.118 & 14 \\
CWE-200 & 0.000 & 0.000 & 0.000 & 9 \\
CWE-502 & 0.333 & 0.143 & 0.200 & 7 \\
CWE-611 & 0.118 & 0.154 & 0.133 & 13 \\
CWE-79 & 0.273 & 0.176 & 0.214 & 17 \\
Negative & 0.442 & 0.633 & 0.521 & 60 \\

\midrule 
\midrule 
accuracy &  & &                         0.375 & 120 \\
macro avg & 0.250 & 0.196 &      0.198 & 120 \\ 
weighted avg & 0.331 & 0.375 &      0.330 & 120 \\
\bottomrule 
\end{tabular}
\smallskip 
\caption{Performance Evaluation of the 'text-davinci-003' Model on 5 CWE Types Using 120 Samples}
\label{tab:gpt_multi}
\end{table}

\begin{table}

\centering
\begin{tabular}{l c c c c}
\toprule
& \multicolumn{4}{c}{Multi} \\ 
\cmidrule(l){2-5} 
CWE type & Precision & Recall &  F1-Score & support \\ 
\midrule 
CWE-20 & 0.118 & 0.118 & 0.116 & 14\\
CWE-200 & 0.081 & 0.080 & 0.079 & 9\\
CWE-502 & 0.057 & 0.056 & 0.055 & 7\\
CWE-611 & 0.108 & 0.107 & 0.105 & 13\\
CWE-79 & 0.138 & 0.140 & 0.137 & 17\\
Negative & 0.500 & 0.500 & 0.499 & 60\\

\midrule 
\midrule 
accuracy &  & &             0.305 & 120 \\
macro avg & 0.167 & 0.167 & 0.165 & 120 \\ 
weighted avg & 0.304 & 0.305 & 0.303 & 120 \\
\bottomrule 
\end{tabular}
\smallskip 
\caption{Performance Evaluation of the Baseline Model on 5 CWE Types Using 120 Samples}
\label{tab:baseline}
\end{table}
To further evaluate the model's capabilities, we selected the top five CWE vulnerability types based on their 
frequency in the dataset (CWE-20, CWE-200, CWE-502, CWE-611, CWE-79). We used this subset of the data to perform multi classification task. 
The results show (Table ~\ref{tab:gpt3_5_multi}, ~\ref{tab:gpt_multi}, ~\ref{tab:baseline}) that the ChatGPT and GPT-3 models achieved an accuracy of 38\%. The tables include precision, recall, F1-score, and support for each of the five CWE types as well as accuracy, macro average, and weighted average metrics.


In the sample, the negative class (representing patched non-vulnerable functions) was predominant, 
and the models exhibited a higher accuracy by predicting this class. However, for a more objective assessment of model's ability to detect vulnerable functions, 
we excluded negative examples from the sample 
(Table ~\ref{tab:gpt3_5_multi_60}, ~\ref{tab:gpt3_multi_60}, ~\ref{tab:baseline_60}). 
Surprisingly, the models' accuracy was lower after excluding the Negative class, indicating that they struggled 
with correctly predicting vulnerable functions.

For the multi-label classification task, we used the following prompt: "Which of these types of vulnerabilities is contained in the code? 
Response in json format: { "type": "CWE-502"} Where instead of "CWE-502" there can be any of the options: CWE-502, CWE-79, CWE-20, CWE-611, 
CWE-200, Negative". We requested the response to be in JSON format for ease of parsing and to prevent unclear model responses. 
The appendix Table  ~\ref{appx:prompts} shows other variants of prompts that we used for this task.

\begin{table} 
\centering
\begin{tabular}{l c c c c}
\toprule
& \multicolumn{4}{c}{Multi} \\ 
\cmidrule(l){2-5} 
CWE type & Precision & Recall &  F1-Score & support \\ 
\midrule 
CWE-20 & 1.000 & 0.071 & 0.133 & 14 \\ 
CWE-200 & 0.000 & 0.000 & 0.000 & 9 \\ 
CWE-502 & 1.000 & 0.143 & 0.250 & 7 \\ 
CWE-611 & 0.250 & 0.077 & 0.118 & 13 \\ 
CWE-79 & 1.000 & 0.176 & 0.300 & 17 \\ 

\midrule 
\midrule 
accuracy & & &  0.100 & 60 \\ 
macro avg & 0.542 & 0.078 & 0.133 & 60 \\ 
weighted avg & 0.688 & 0.100 & 0.171 & 60 \\ 
\bottomrule 
\end{tabular}
\smallskip 
\caption{Performance Evaluation of the 'gpt-3.5-turbo' Model on 5 CWE Types Using 60 Samples} 
\label{tab:gpt3_5_multi_60}
\end{table}

\begin{table} 
\centering
\begin{tabular}{l c c c c}
\toprule
& \multicolumn{4}{c}{Multi} \\ 
\cmidrule(l){2-5} 
CWE type & Precision & Recall &  F1-Score & support \\ 
\midrule 
CWE-20 & 1.000 & 0.071 & 0.133 & 14 \\
CWE-200 & 0.000 & 0.000 & 0.000 & 9 \\
CWE-502 & 1.000 & 0.143 & 0.250 & 7 \\
CWE-611 & 0.286 & 0.154 & 0.200 & 13 \\
CWE-79 & 1.000 & 0.176 & 0.300 & 17 \\
\midrule 
\midrule 
accuracy & & & 0.117 & 60 \\
macro avg & 0.548 & 0.091 & 0.147 & 60 \\
weighted avg & 0.695 & 0.117 & 0.189 & 60 \\
\bottomrule 
\end{tabular}
\smallskip 
\caption{Performance Evaluation of the 'text-davinci-003' Model on 5 CWE Types Using 60 Samples}
\label{tab:gpt3_multi_60}
\end{table}

\begin{table} 
\centering
\begin{tabular}{l c c c c}
\toprule
& \multicolumn{4}{c}{Multi} \\ 
\cmidrule(l){2-5} 
CWE type & Precision & Recall &  F1-Score & support \\ 
\midrule 
CWE-20 & 0.238 & 0.239 & 0.236 & 14 \\
CWE-200 & 0.150 & 0.147 & 0.145 & 9 \\
CWE-502 & 0.112 & 0.112 & 0.109 & 7 \\
CWE-611 & 0.215 & 0.214 & 0.211 & 13 \\
CWE-79 & 0.282 & 0.283 & 0.280 & 17 \\
\midrule 
\midrule 
accuracy & & & 0.218 & 60 \\
macro avg & 0.199 & 0.199 & 0.196 & 60 \\
weighted avg & 0.218 & 0.218 & 0.214 & 60 \\
\bottomrule 
\end{tabular}
\smallskip 
\caption{Performance Evaluation of the Dummy Classifier on 5 CWE Types Using 60 Samples} 
\label{tab:baseline_60}
\end{table}

\begin{table} 
\centering
\begin{tabular}{l c c c c }
\toprule
 &  \rotatebox{90}{gpt-3.5(0)} & \rotatebox{90}{gpt-3.5(1)} & \rotatebox{90}{gpt-3.5(2)} & \rotatebox{90}{gpt-3.5(3)}\\ 
\midrule 
$g_1$        & 7     & 6     & 7     & 6\\
\hline
$g_2$     & 14    & 14    & 12    & 12\\


\bottomrule 
\end{tabular}
\smallskip 
\caption{Accuracy difference for vulnerable and non-vulnerable groups. 
\\ Where $g_1$ - Accuracy across vulnerable functions (60 examples), $g_2$ - Accuracy across non-vulnerable conterparts (60 examples) } 
\label{tab:more}
\end{table}

We observed an interesting anomaly: the ChatGPT predicts a vulnerability type more accurately for fixed functions.
Another words, if we assigned the same labels to fixed functions as it was for its vulnerable countepart, the ChatGPT would predict these labels about two times more
accurate. Table ~\ref{tab:more} shows accuracy for two groups across experiments with different prompts.

The accuracy difference could be explained if the model being trained on code with already fixed functions, 
or it could be a random effect due to the small sample size.


\section{Threats to validity}
\subsection{Conclusion validity}
The conclusion validity of the study may be threatened by the small sample size used in the experiment, which may limit the statistical power of the analysis. 

\subsection{Construct validity}
The construct validity of the study may be threatened by the use of a single GPT model to detect vulnerabilities in Java code. The model may have inherent limitations and biases that may affect its ability to accurately detect vulnerabilities, and the results may not be generalizable to other models or approaches to vulnerability detection.

One more of the factors that may affect the validity of the study is the choice of prompt for binary and multi-label classification. For example, in our case, for multi-label classification, we used prompt that asked the model about a specific type of vulnerability. However, there are other ways to classify vulnerabilities. The model may not be well integrated into these types of vulnerabilities. Perhaps a more careful selection of prompts could give different results.

Another factor that may influence the construct validity is the model's ability to generate a chain of thought (\cite{wei2022chain}), which refers to a series of intermediate reasoning steps. Generating a chain of thought has been shown to significantly improve the reasoning capabilities of large language models. In our study, we did not explicitly explore the chain-of-thought prompting technique, which involves providing a few chain of thought demonstrations as exemplars in prompting. This approach might have improved the model's ability to detect vulnerabilities more accurately.

\subsection{Internal validity}
This study's internal validity may be threatened by the possibility that the GPT model was trained on examples that contained instances of CWE vulnerabilities but not on data that demonstrated how vulnerabilities were remediated. Further training may be necessary to improve the model's ability to accurately distinguish between vulnerable and remediated functions.

Another factor that may threaten the internal validity of this study is the manual selection of vulnerable code by domain experts. Although domain experts can identify known vulnerabilities, there may be instances where vulnerabilities are missed or incorrectly labeled. This could affect the accuracy of the dataset and the subsequent evaluation of the GPT model's performance. Furthermore, as the dataset is compiled based on domain expert opinions, it may introduce some bias, which could also impact the results. To address this threat, future studies could consider using multiple domain experts to review and label the data, employ automated tools for vulnerability identification, or use larger and more diverse datasets.

The subset of files extracted from the dataset for our experiments was chosen based on a specific criterion: only files with a single modified function were included. This may introduce a selection bias, as the distribution of vulnerabilities in this subset may not reflect the actual distribution of vulnerabilities in the real world. Therefore, the generalizability of our results to other datasets and scenarios may be limited. This represents a potential internal validity threat to our study.

\subsection{External validity}
The use of a single dataset may limit the generalizability of the findings to other datasets.
\balance
\section{Conclusion}

In conclusion, our experiments with GPT-3 and ChatGPT models indicate that their current capabilities for effectively detecting vulnerabilities in code are limited. Although they have been trained on vast amounts of data, these models did not achieve high performance on our test data. While natural language processing models have demonstrated impressive results in numerous areas, it is evident that their application in vulnerability detection tasks requires further refinement and investigation.

As a possible direction for future research, it may be worth exploring the impact of the chain-of-thought prompting technique on the performance of the ChatGPT and GPT-3 models in vulnerability detection tasks. This technique has been shown to improve the reasoning abilities of large language models (\cite{wei2022chain}), and our study did not explicitly investigate it. By providing a few chain-of-thought demonstrations as exemplars in prompting, we might improve the models' ability to detect vulnerabilities more accurately.

We believe there is potential for these models to improve and contribute to vulnerability detection in the future, with more targeted research and advancements in the field. It would be worthwhile to explore the efficacy of upcoming large language models, such as GPT-4, in addressing this task and to identify ways to enhance their performance in this specific domain.


\bibliographystyle{ACM-Reference-Format} 
\bibliography{references}

\clearpage

\appendix
\renewcommand{\thetable}{\Alph{table}}
\setcounter{table}{0}

\begin{table}[h!]
\centering
\begin{tabular}{@{}rl@{}|@{\hspace{2mm}}rl@{}}
\toprule
CWE & Count \hspace{2mm} &  CWE & Count \hspace{2mm} \\
\midrule
CWE-538 & 1 & CWE-400 & 4 \\
CWE-79 & 17 & CWE-862 & 5 \\
CWE-552 & 1 & CWE-352 & 2 \\
NVD-CWE-Other & 10 & CWE-787 & 2 \\
CWE-74 & 2 & CWE-476 & 1 \\
NVD-CWE-noinfo & 9 & CWE-428 & 1 \\
CWE-264 & 11 & CWE-200 & 9 \\
CWE-345 & 2 & CWE-863 & 3 \\
CWE-20 & 14 & CWE-915 & 1 \\
CWE-22 & 11 & CWE-388 & 1 \\
CWE-611 & 13 & CWE-203 & 1 \\
CWE-732 & 1 & CWE-404 & 2 \\
CWE-502 & 7 & CWE-284 & 1 \\
CWE-269 & 1 & CWE-384 & 1 \\
CWE-362 & 1 & CWE-668 & 2 \\
CWE-601 & 1 & CWE-240 & 1 \\
CWE-287 & 4 & CWE-399 & 1 \\
CWE-401 & 1 & CWE-335 & 1 \\
CWE-835 & 5 & CWE-754 & 1 \\
CWE-310 & 1 & CWE-755 & 1 \\
\end{tabular}
\caption{Number of different CWE}
\label{appx:distribution_CWE}
\end{table}

\begin{table}[]
\centering
\begin{tabular}{|l|c|}
\hline
\multirow{2}{*}{\textbf{Project}} & \multicolumn{1}{c|}{\textbf{Number of}} \\
 & \multicolumn{1}{c|}{\textbf{Vulnerabilities}} \\ \hline
apache--tomcat & 32 \\
jenkinsci--jenkins & 21 \\
apache--tomcat80 & 5 \\
FasterXML--jackson-databind & 4 \\
apache--activemq & 4 \\
theonedev--onedev & 2 \\
apache--lucene-solr & 2 \\
javamelody--javamelody & 2 \\
opennetworkinglab--onos & 2 \\
pippo-java--pippo & 2 \\
apache--cordova-android & 2 \\
apache--tomcat55 & 2 \\
traccar--traccar & 2 \\
undertow-io--undertow & 2 \\
apache--struts & 2 \\
apache--cxf & 2 \\
apache--flink & 1 \\
apache--pdfbox & 1 \\
OWASP--json-sanitizer & 1 \\
... & \\
 \hline
\end{tabular}
\caption{Projects and Number of Vulnerabilities Used in the Dataset}
\label{appx:project_count}
\end{table}

\begin{table}[h!]
  \begin{center}
    
    \begin{tabular}{|l|}
    \hline
      \textbf{Prompt}\\
      \hline
   "Which of these types of vulnerabilities is contained in the code?  \\
1. CWE-502: Deserialization of Untrusted Data \\
2. CWE-79: Cross-site Scripting \\
3. CWE-20: Improper Input Validation \\
4. CWE-611: Improper Restriction of XML External Entity Reference \\
5. CWE-200: Exposure of Sensitive Information to an Unauthorized Actor \\
6. There is no vulnerability in the code"\\
\hline
     "What type of vulnerability is in the code. Choose only one from the \\
     following options.  \\
1. CWE-502: Deserialization of Untrusted Data  \\
2. CWE-79: Cross-site Scripting  \\
3. CWE-20: Improper Input Validation  \\
4. CWE-611: Improper Restriction of XML External Entity Reference  \\
5. CWE-200: Exposure of Sensitive Information to an Unauthorized Actor  \\
6. There is no vulnerability in the code"\\
\hline
     "Which of these types of vulnerabilities is contained in the code?   \\
1. CWE-502: Deserialization of Untrusted Data  \\
2. CWE-79: Cross-site Scripting  \\
3. CWE-20: Improper Input Validation  \\
4. CWE-611: Improper Restriction of XML External Entity Reference  \\
5. CWE-200: Exposure of Sensitive Information to an Unauthorized Actor"\\
\hline
     "Which of these types of vulnerabilities is contained in the code?   \\
Response in json format: { "type": "CWE-502"}   \\ 
Where instead of "CWE-502" there can be any of the options: \\
CWE-502, CWE-79, CWE-20, CWE-611, CWE-200, Negative"\\
\hline
    \end{tabular}
    
  \end{center}
  \caption{Multi-label Prompts}
  \label{appx:prompts}
\end{table}

\end{document}